\documentclass[a4paper, 10pt]{article}

\usepackage[style=apa, 
backend=biber, 
uniquename=false,
url=true, 
doi=true]{biblatex}
\usepackage{hyperref}
\usepackage[dvipsnames]{xcolor}
\usepackage{colortbl}
\hypersetup{
plainpages=false,%
bookmarksopen=true,%
bookmarksnumbered=false,%
bookmarksdepth=5,
breaklinks=true,
colorlinks=true,
citecolor=BlueViolet,
linkcolor=BlueViolet,
urlcolor=blue,
}
\usepackage{citeauthor}

\usepackage{mathrsfs} 

\usepackage[T1]{fontenc}
\usepackage[utf8]{inputenc}

\usepackage{times} 

\usepackage{bm}
\usepackage{amsmath}
\usepackage{amssymb}

\usepackage{SpringerPsychometrika}

\usepackage{fancyhdr}

\fancypagestyle{firstpage}{
\fancyhf{}
\rhead{\textit{arXiv Preprint}}
\lhead{}
\chead{}
\cfoot{}
\lfoot{Prepared for submission}

}

\fancypagestyle{fund_footer}{%
\fancyhf{}

\fancyfoot[L]{\small}
}

\usepackage{setspace}

\usepackage[noblocks, affil-it]{authblk}

\usepackage{abstract}
\renewenvironment{abstract}{\section*{\centering\abstractname}}{}
\renewcommand{\abstractname}{}

\usepackage{graphicx}
\newcommand{\orcid}[1]{\href{#1}{\includegraphics[width=10pt]{orcid.png}}}

\usepackage{graphicx}

\usepackage{booktabs}
\usepackage{multicol}
\usepackage{multirow}
\usepackage{dcolumn}
\usepackage{dcolumn}
\usepackage{lscape}

\newcolumntype{L}{>{$}l<{$}} 
\newcolumntype{C}{>{$}c<{$}} 
\newcolumntype{R}{>{$}r<{$}} 

\usepackage{float}

\usepackage{threeparttable}
\usepackage{adjustbox}
\usepackage{algorithm}
\usepackage{algorithmic}

\begin{document}

\title{\vspace{-1.5\bigskipamount} \normalsize \uppercase{Assessing the Performance of Diagnostic Classification Models in Small Sample Contexts with Different Estimation Methods}}
\author[ ]{\fontsize{11pt}{11pt}{\scshape Motonori Oka}}
\author[ ]{\fontsize{11pt}{11pt}{\scshape Kensuke Okada}}
\affil[ ]{\small{GRADUATE SCHOOL OF EDUCATION, THE UNIVERSITY OF TOKYO}}
\date{}
\maketitle

\thispagestyle{firstpage}

\vspace{-2.7\bigskipamount}
\begin{abstract}
\begin{center}\footnotesize
\begin{minipage}{\dimexpr\paperwidth-79mm}
\setlength{\parindent}{12pt}
Fueled by the call for formative assessments, diagnostic classification models (DCMs) have recently gained popularity in psychometrics. Despite their potential for providing diagnostic information that aids in classroom instruction and students' learning, empirical applications of DCMs to classroom assessments have been highly limited. This is partly because how DCMs with different estimation methods perform in small sample contexts is not yet well-explored. Hence, this study aims to investigate the performance of respondent classification and item parameter estimation with a comprehensive simulation design that resembles classroom assessments using different estimation methods. The key findings are the following: (1) although the marked difference in respondent classification accuracy was not observed among the maximum likelihood (ML), Bayesian, and nonparametric methods, the Bayesian method provided slightly more accurate respondent classification in parsimonious DCMs than the ML method, and in complex DCMs, the ML method yielded the slightly better result than the Bayesian method; (2) while item parameter recovery was poor in both Bayesian and ML methods, the Bayesian method exhibited unstable slip values owing to the multimodality of their posteriors under complex DCMs, and the ML method produced irregular estimates that appear to be well-estimated due to a boundary problem under parsimonious DCMs. \\

\end{minipage}
\end{center}
\end{abstract}

\vspace{0.9\bigskipamount}
\section{Introduction}\label{intro}
Diagnostic classification models (DCMs), also referred to as cognitive diagnostic models, have received considerable attention from methodological and applied researchers for their potential use in formative assessments. These models offer a novel means of analyzing test scores. In contrast to a traditional approach in which item response theory commonly assumes a unidimensional latent ability for locating examinees on a continuum, DCMs aim to classify students according to their mastery state of multiple skills and abilities, which are quantified as discrete latent variables called \emph{attributes}. The goal of DCMs then is to assign the students to one of the possible combinations of mastery or non-mastery of the attributes, which are called \emph{attribute mastery profiles}. Consider the math test as an example. Teachers may be interested to know which concepts of mathematics students struggle to learn. While the traditional approach generally provides students' scores, which indicate their position on the continuum of overall mathematical ability, DCMs enable teachers to inspect students' mastery profiles of fine-grained attributes for mathematics \parencite{henson_test_2005}. By utilizing this detailed knowledge extracted from DCMs, teachers can prepare a teaching plan that provides individualized classroom instruction. Students can also receive diagnostic information on the strengths and weaknesses in their learning. 

Owing to these advantages of DCMs, many applied researchers have used diagnostic classification modeling in such domain fields as education and clinical psychology \parencite[e.g.,][]{ranjbaran_developing_2017,jang_improving_2019,wang_development_2019}. However, most DCM applications deal with large sample sizes. DCM applications to small sample sizes have been limited, which does not align with the original prospect of adopting DCMs to enhance classroom activities \parencite{sessoms_applications_2018}. Despite the general scarcity of applications to small sample sizes, some small-scale field studies supported the sizable utility of DCMs \parencite{li_latent_2016,jang_how_2015,jang_validity_2005}. For instance, \textcite{jang_validity_2005} applied one of the DCMs, the fusion model, to the assessment data for reading skills collected from 27 students who enrolled in two TOEFL test preparation courses. The resulting report stated that the students considered the received diagnostic information helpful to recognize what they did not understand, and the teachers involved with the two preparation courses regarded the diagnostic information as useful for understanding students' strengths and weaknesses in reading skills. As this study indicates, DCMs hold vast potential to obtain diagnostic information that enriches classroom instruction as well as students' learning.

Although the dissemination of such applications will fulfill the original promise of DCMs to aid in students' learning via classroom assessments, it is still a pending question how accurately DCMs' parameters can be estimated using classroom assessment data that typically involve small sample sizes. Regarding simulation studies pertinent to small sample sizes, \textcite{basokcu_classi_2014} considered 30, 50, 100, 200, and 400 respondents and Q-matrices---which specify item and attribute relationships---yielding different model-fit values in the simulation design to assess the performance of the deterministic inputs, noisy ``and" gate \parencite[DINA;][]{junker_cognitive_2001} model and generalized DINA \parencite[GDINA;][]{delatorre_generalized_2011} model with respect to fit indices and classification accuracy. This study shows that the decrease in sample sizes moderately reduced classification accuracy, although the effect of misfitted Q-matrices was predominant. \textcite{paulsen_examining_2019} conducted a simulation study to investigate the performance of the DINA model using maximum likelihood (ML) estimation, a nonparametric method \parencite{chiu_nonparametric_2013}, and a supervised neural network \parencite{cui_statistical_2016} in small sample sizes (i.e. 25, 50, 150, and 1000). The study's findings revealed that when item discrimination, sample size, and the number of items increased, the estimation accuracy for attributes improved, and when the number of attributes and the degree of the misspecification for a Q-matrix were large, the estimation accuracy deteriorated. \textcite{yang_improved_2019} investigated how fit indices performed in small sample sizes (i.e. 50, 75, 100, and 200) and confirmed that Akaike information criterion (AIC) is an appropriate choice for model selection under small sample conditions. Finally, \textcite{sen_sample_2021} examined the classification accuracy and item parameter estimation of four DCMs using the ML method under the sample sizes of 50 to 5000 and found that the increase in the number of attributes and decrease in the number of respondents negatively contributed to item parameter recovery. They also reported that parsimonious DCMs performed better in terms of classification accuracy and item parameter recovery than complex DCMs in small sample sizes.

These studies help understand the behavior of the DCMs' estimates under small sample conditions; however, the focus of their simulation designs is limited when it comes to small-sample applications. First, it is common for real classroom assessment data to contain a few items with irregular responses known as a \emph{perfect response pattern} \parencite{levy_bayesian_2016}, where all the respondents give correct, or incorrect, responses. When this happens, the ML method cannot provide estimates of model parameters because the likelihood tends to infinity. However, the aforementioned simulation studies except for \textcite{paulsen_examining_2019} only considered the ML method that can be infeasible in small sample sizes. Hence, other estimation methods that can cope with such irregular responses, such as Bayesian estimation, should be considered. Second, although the number of students in one classroom can be less than 50 in many cases, the previous studies tend to limit a range of sample sizes starting from 50. This limitation underscores the distinctive importance of a simulation study focusing on more realistic sample sizes as a classroom assessment. Thus, a far-reaching simulation study with an emphasis on classroom sample sizes is required.

Therefore, the present study aims to conduct a simulation study with a comprehensive design that targets classroom sample sizes and encompasses all the frequently used estimation methods, namely, ML, Bayesian, and nonparametric estimations, as well as popular DCMs, namely, the DINA model, deterministic inputs, noisy ``or" gate \parencite[DINO;][]{templin_measurement_2006} model, reduced reparameterized unified model \parencite[RRUM;][]{hartz_fusion_2008}, compensatory RUM \parencite[CRUM;][]{rupp_diagnostic_2010}, and logliner cognitive diagnostic model \parencite[LCDM;][]{henson_defining_2009}. For this simulation, seven manipulated factors that resemble typical classroom assessments were considered. We then investigate the performance of respondent classification and item parameter estimation under various conditions.

The remainder of the paper is organized as follows. In section 2, we introduce the five DCMs employed in the simulation study and provide explanations for the manipulated factors in the simulation design, the measures of accuracy, and the settings for estimation. In section 3, the results of the simulation are presented. In the final section, we discuss the limitations of this study and future research directions. 

\section{Methods}
\subsection{Diagnostic Classification Models}

In this paper, the subscripts $i\;(1, \ldots, N)$, $j\;(1, \ldots, J)$, and $k\;(1, \ldots, K)$ are used to express respondents, items, and attributes, respectively. In addition, let $\mathbf{X}$ be the $N \times K$ response matrix, whose element $x_{ij}$ takes the value of 1 when respondent $i$ gives a correct answer to item $j$, and 0 otherwise.
 
DCMs formulate the probability of a correct response to item $j$ based on the attribute mastery profile matrix $\mathbf{A}$ and Q-matrix $\mathbf{Q}$. $\bm{\alpha}_i=(\alpha_{i1}, \ldots, \alpha_{iK})$ is the $i$-th row of the matrix $\mathbf{A}$ and represents the mastery state of $K$ attributes for respondent $i$. Generally, the possible number of attribute mastery profiles amount to $L=2^K$. Accordingly, respondents are supposed to have one of the $L$ attribute mastery profiles. A Q-matrix specifies which attributes are required to answer item $j$ correctly and is usually constructed by domain-experts before implementing an assessment. The element of a Q-matrix is denoted by $q_{jk}=\{0, 1\}$. When $q_{jk}$ equals 1, attribute $k$ is 
required for the correct answer to item $j$ and vice versa. 

In the following subsections, we introduce the five DCMs considered in this simulation study. 

\subsubsection*{(1) The DINA model}

The DINA model holds the noncompensatory assumption in which a respondent needs to acquire all the required attributes for item $j$ to answer it correctly. This assumption is embedded in the ideal response of the DINA model:
\begin{align*}
\eta_{ij} = \prod_{k=1}^K \alpha_{ik}^{q_{jk}},
\end{align*}
where $\eta_{ij}$ equals 1 when the attribute mastery profile $\bm{\alpha}_i$ satisfies $\alpha_{ik} \geq q_{jk}$ for all $k$. The item response function of the DINA model is then formulated as
\begin{align*}
& p(x_{ij}=1 \vert s_j, g_j, \bm{\alpha}_i, \bm{q}_{j}) = \Bigl( (1 - s_j)^{x_{ij}}s_j^{1-x_{ij}} \Bigl)^{\eta_{ij}} \Bigl( g_j^{x_{ij}}(1-g_j)^{1-x_{ij}} \Bigl)^{1 - \eta_{ij}}.
\end{align*}
The slip parameter $s_j=p(x_{ij}=0 \vert \eta_{ij}=1)$ and the guessing parameter $g_j=p(x_{ij}=1 \vert \eta_{ij}=0)$ represent the probability of giving a wrong answer to item $j$ when the ideal response equals 1 and the probability of giving a correct answer to item $j$ when the ideal response equals 0, respectively. 

\subsubsection*{(2) The DINO model}

In contrast to the DINA model, the DINO model assumes the compensatory nature in its item response, where a respondent can obtain a correct answer to item $j$ with the mastery of at least one among the specified attributes for that item. The ideal response of the DINO model reflects this compensatory assumption. It is defined as 
\begin{align*}
\omega_{ij} = 1 - \prod_{k=1}^K (1 - \alpha_{ik})^{q_{jk}},
\end{align*}
where $\omega_{ij}$ equals 1 when respondent $i$ masters at least one of the specified attributes for item $j$. The item response function is formulated as follows:
\begin{align*}
& p(x_{ij}=1 \vert s_j, g_j, \bm{\alpha}_i, \bm{q}_{j}) =  \Bigl( (1 - s_j)^{x_{ij}}s_j^{1-x_{ij}} \Bigl)^{\omega_{ij}} \Bigl( g_j^{x_{ij}}(1-g_j)^{1-x_{ij}} \Bigl)^{1 - \omega_{ij}}.
\end{align*}
The slip and guessing parameters correspond to $s_j=p(x_{ij}=0 \vert \omega_{ij}=1)$ and $g_j=p(x_{ij}=1 \vert \omega_{ij}=0)$, respectively. 

\subsubsection*{(3) The RRUM}

The DINA model does not differentiate the correct-response probabilities of item $j$ among all the respondents with the attribute mastery profiles that do not suffice $\alpha_{ik} \geq q_{jk}$ for all $k$. However, it may be more natural to consider that these correct-response probabilities are different across respondents who acquire none of the specified attributes and those who lack only one of them. To flexibly address this type of case, the RRUM is formulated with the baseline parameter $\pi_j$, which is the probability of correct responses to item $j$, given the mastery of all the specified attributes, and the penalty parameter $r_{jk}$ that reduces $\pi_j$ when attribute $k$ is not acquired \parencite{zhan_using_2019}. The item response function of the RRUM is defined as
\begin{align*}
p(x_{ij}=1 \vert \pi_j, r_{j1}, \ldots, r_{jK}, \bm{\alpha}_i, \bm{q}_{j}) = \pi_j \prod_{k=1}^K r_{jk}^{(1 - \alpha_{ik}){q_{jk}}}.
\end{align*}
It should be noted that the slip and guessing values of the RRUM for item $j$ are equivalent to $1-\pi_j$ and $\pi_j \prod_{k=1}^K r_{jk}$, respectively. 

\subsubsection*{(4) The CRUM}

Being analogous to the RRUM, the CRUM is developed to allow for flexible modeling on the correct-response probabilities for the respondents who master at least one of the specified attributes and those who master more than one. The CRUM consists of two components: the intercept parameter $\lambda_{j0}$ and slope parameters $\lambda_{jk}$. As with the RRUM, these parameters are defined at item levels. Since the CRUM is in the form of logistic regression with attributes, the slope parameters can be interpreted as main-effect terms of attributes that increase the logit of the probability of a correct response \parencite{rupp_diagnostic_2010}. The item response function of the CRUM expressed in logit form is given as 
\begin{align*}
\mathrm{logit}\Bigl( p(x_{ij}=1 \vert \bm{\lambda_{j}}, \bm{\alpha}_i, \bm{q}_{j}) \Bigl) = \lambda_{j0} + \sum_{k=1}^K \lambda_{jk}\alpha_{ik}q_{jk}.
\end{align*}
$1 - \mathrm{logit}^{-1}\Bigl(\lambda_{j0} + \sum_{k=1}^K \lambda_{jk}q_{jk} \Bigl)$ and $\mathrm{logit}^{-1}\Bigl(\lambda_{j0} \Bigl)$ represent the slip and guessing values for item $j$, respectively. 

\subsubsection*{(5) The LCDM}

The LCDM is the generalization of the DINA model, DINO model, RRUM, and CRUM in the sense that these models can be derived by imposing constraints on the LCDM's parameters. The saturated form of the LCDM allows all the interaction terms of attributes. Hence, this model can flexibly account for any combinations of effects from the attributes. The item response function of the LCDM expressed in logit form is defined as
\begin{align*}
& \mathrm{logit}\Bigl( p(x_{ij}=1 \vert \bm{\lambda_{j}}, \bm{\alpha}_i, \bm{q}_{j}) \Bigl)\\
&\quad =
\lambda_{j0} + \sum_{k=1}^K \lambda_{jk}\alpha_{ik}q_{jk} + \sum_{k=1}\sum_{k'>1}^K \lambda_{jk, (k,k')}\alpha_{ik}\alpha_{ik'}q_{jk}q_{jk'} + \cdots.
\end{align*}
The interaction effects after the second term of the right-hand side of the above equation represent the summation of all the interaction terms for attributes. The slip and guessing values of the LCDM for item $j$ correspond to $1 - \mathrm{logit}^{-1}\Bigl(\lambda_{j0} + \sum_{k=1}^K \lambda_{jk} + \sum_{k=1}\sum_{k'>1}^K \lambda_{jk, (k,k')}q_{jk}q_{jk'} + \cdots \Bigl)$ and $\mathrm{logit}^{-1}\Bigl(\lambda_{j0} \Bigl)$, respectively. 

It should be noted that the LCDM is equivalent to the GDINA model with the logit link function. 

\subsection{Simulation Design}

In the current simulation study, we considered seven manipulated factors to mimic the conditions under which typical classroom assessment data are collected. The factors' specifications are partly based on the previous simulation study \parencite{paulsen_examining_2019} and partly informed by settings of actual assessments. The overall simulation design is summarized in Table \ref{tab:Table1}. 

\begin{table}[!htbp]
\centering
\caption{Summary of simulation design}
\resizebox{\linewidth}{!}{
\begin{tabular}{lcl}
\toprule
Design Factor & \multicolumn{1}{l}{Number of Levels} & Values of Levels \\
\midrule
Sample Size & 3 & $N=20, 40, 160$ \\
Number of Items & 2 & $J=20, 40$ \\
Number of Attributes & 2 & $K=4,5$ \\
Item Discrimination & 2 & \multicolumn{1}{l}{\begin{tabular}{l}\hspace{-8.5pt} High: $s_j, g_j \sim \mathrm{Uniform}(0, 0.15)$, \\ \hspace{-8.5pt} Low: $s_j, g_j \sim \mathrm{Uniform}(0.25, 0.4)$ \end{tabular}} \\
Q-matrix Misspecification & 3 & 0\%, 10\%, 20\% \\
Estimation Methods & 3 & ML, Bayes, nonparametric \\
\midrule
Models & 5 & DINA, DINO, RRUM, CRUM, LCDM \\
\bottomrule 
\end{tabular}%
}
 \begin{tablenotes}
 \item \footnotesize{\textit{Note.} The factors of the number of items and attributes construct the factor of a Q-matrix with four levels.}
 \end{tablenotes}
\label{tab:Table1}%
\end{table}

\subsubsection*{(1) Sample Size}

The first design factor is the sample size, which we set to 20, 40, and 160. These levels are intended to mimic average classroom and grade sizes. As \textcite{sun_designing_2013} illustrated in the application of a DCM for fraction problems to 144 sixth grade students in an elementary school, it is reasonable to include the level of 160 in this factor as the possible sample size for a classroom assessment. 

\subsubsection*{(2) The Number of Items}

The second design factor is the number of items. The levels of this factor were assigned 20 and 40 items. Since this simulation study approximates conditions of classroom assessments, we regard 40 as the maximum number of items that can be implemented within one period of classes because of the number of items in the math test of the National Center Test for University Admissions containing about 40 items to be solved in an hour \parencite{national_center_for_university_entrance_examinations_university_2020}. 

\subsubsection*{(3) The Number of Attributes}

The third design factor is the number of attributes. The levels of this factor were set to four and five. Although large-scale assessments are often associated with more than eight dimensionalities of attributes \parencite{sessoms_applications_2018}, a classroom assessment typically does not have such high dimensionalities. This is because including sufficient items to capture all the combinations of attribute mastery profiles leads to infeasible testing time in one period of classes. Thus, we consider four and five attributes as plausible dimensionalities of attributes for a classroom assessment. 

\subsubsection*{(4) Q-matrix}

The fourth design factor is the Q-matrix. Because the levels for the number of items and attributes are both two, the number of possible Q-matrices becomes four. The four-attribute Q-matrix with 20 items includes 15 $(=2^4-1)$ items measuring all possible combinations of attributes and five randomly selected items from the set of possible items that require two and three attributes. For the four-attribute Q-matrix with 40 items, two sets of 15 items measuring all possible combinations of attributes and ten randomly selected items from the same set of items measuring two and three attributes are included. The five-attribute Q-matrix with 20 items contains all the possible (i.e., five) items that measure one attribute and 15 randomly selected items from the set of possible items that require two and three attributes. Subsequently, the five-attribute Q-matrix with 40 items contains two sets of all the possible items that measure one attribute and 30 randomly selected items from the set of possible items that require two and three attributes. We confirmed that the Q-matrices for 20 items satisfy completeness and some of the identification conditions for DCMs' parameters, and the Q-matrices for 40 items satisfy all the identification conditions for them. These conditions are necessary for the identification of DCMs' parameters \parencite{xu_identifying_2018,liu_constrained_2020}. 

\subsubsection*{(5) Q-matrix Misspecification}

The true Q-matrix is not always correctly specified by domain experts. On the contrary, Q-matrix misspecification frequently occurs owing to the inadequacy of underlying knowledge, which leads to the deterioration of respondent classification and item parameter estimation accuracy \parencite{rupp_effects_2008,kunina-habenicht_impact_2012}. Hence, the effect of Q-matrix misspecification has been examined in DCM literature. In this study, we set 0\%, 10\%, and 20\% misspecification for the entries of a Q-matrix. The misspecified entries of a Q-matrix were randomly selected and changed in a manner such that $q^{true}_{jk}=1 \rightarrow q^{miss}_{jk}=0$ or $q^{true}_{jk}=0 \rightarrow q^{miss}_{jk}=1$. The former and latter cases of misspecification are referred to as underspecification and overspecification, respectively. To balance the occurrence of these two types of misspecification, we set the number of underspecified and overspecified elements to be equal. Additionally, ten different randomly-misspecified Q-matrices were generated for each correct Q-matrix, and each of them was applied for every ten replications. This approach was proposed in \textcite{kunina-habenicht_impact_2012} to generalize the impact of Q-matrix misspecification by randomly changing the patterns of misspecification. It should be noted that the items measuring a single attribute were excluded from the random selection process to ensure the identification conditions of a Q-matrix, and we confirmed that all the misspecified Q-matrices satisfy the same conditions of completeness and identifiability as the corresponding true Q-matrices. 

\subsubsection*{(6) Item Discrimination}

The sixth design factor is item discrimination. Item discrimination in DCMs can be captured by the slip and guessing values \parencite{de_la_torre_empirically_2008} because an item with high slip or guessing values, which corresponds to low discriminating power, indicates that respondents are likely to give an aberrant response that is inconsistent with the ideal response for that item. Accordingly, the items with high discriminating power are expected to accurately differentiate the mastery or non-mastery of the necessary attributes because their actual responses align with the corresponding ideal responses. In the simulation study, we employed two levels of item discrimination: ``high" and ``low." For the ``high" item discrimination, true slip and guessing values were sampled from the uniform distribution $\mathrm{Uniform}(0, 0.15)$. Similarly, for the ``low" item discrimination, they were sampled from $\mathrm{Uniform}(0.25, 0.4)$. As items in a classroom assessment may not be as well-developed as a high-stakes test in a large-scale assessment, the low item discrimination approximates that type of situation in test development. 

\subsubsection*{(7) Estimation Methods} 

The final design factor is estimation methods. Since the data from a classroom assessment often include certain items with a perfect response pattern that the ML method cannot manage, we additionally consider a Bayesian method for respondent classification and item parameter estimation and a nonparametric method for respondent classification, both of which can perform an estimation given the data involving perfect response patterns. The settings for these estimation methods are explained in section 2.5. 

\subsection{Data Generation}

To generate attribute mastery profiles, the multivariate normal threshold model \parencite{chiu_nonparametric_2013} was adopted. The discrete value of $\alpha_{ik}$ pertains to an underlying random variable $\theta_{ik}$ following the multivariate normal distribution $\theta_{ik} \sim \mathrm{MVN}(\bm{0}_K, \bm{{\Sigma}})$. Here, the covariance matrix was constructed as follows:
\begin{align*}
\bm{{\Sigma}} = \left( \begin{array}{ccc} 1 & \cdots & \rho \\ \vdots & \ddots & \vdots \\ \rho & \cdots & 1 \end{array} \right).
\end{align*}
This means that the off-diagonal entries of $\bm{{\Sigma}}$ are the correlation coefficient $\rho$ which is shared for each pair of attributes. In this study, we followed the same specification of \textcite{chiu_nonparametric_2013} in which the correlation coefficient $\rho$ was set to 0.5. The discrete value of $\alpha_{ik}$ was obtained by applying the following criteria:
\begin{align}
\alpha_{ik} = \begin{cases}
1 \;\mathrm{if}\; \theta_{ik} \geq \phi^{-1}\left(\frac{k}{K+1}\right) & \\
 0 \;\mathrm{otherwise} &
\end{cases},
\end{align}
where $\phi^{-1}(\cdot)$ is the inverse function of the cumulative density function for the standard normal distribution. Eq. (1) assumes that each attribute has different difficulty levels for its mastery.

Next, one hundred artificial datasets were generated under each condition by using the ``simGDINA" function in the GDINA package \parencite{ma_gdina_2020} for the R statistical software \parencite{r_development_core_team_r_2020}. Given the specifications for a Q-matrix, true values of the slip and guessing values, and generating assumption on attributes, this function generates artificial response data. The slip and guessing parameters are not assigned directly to the item parameters of the RRUM, CRUM, and LCDM. However, the true values for their item parameters were randomly generated in a manner such that the values of the slip and guessing parameters recalculated with the generated item parameters of these three models become equivalent to the true slip and guessing values specified in the simGDINA function. This random generation expands the generalizability of the specifications for the true values of their item parameters. 

\subsection{Measure of Accuracy}

The accuracy of respondent classification was assessed by the element-wise attribute classification rate (EACR) and the pattern-wise attribute classification rate (PACR). They are given as
\begin{align*}
EACR_{\alpha_k} &= \frac{1}{100} \sum_{m=1}^{100} \frac{1}{N} \sum_{i=1}^N I(\hat{\alpha}_{mik} = \alpha^{true}_{mik}),\\
PACR_{\bm{\alpha}} &= \frac{1}{100} \sum_{m=1}^{100} \frac{1}{N} \sum_{i=1}^N I(\hat{\bm{\alpha}}_{mi} = \bm{\alpha}^{true}_{mi}),
\end{align*}
where $I(\cdot)$ is the indicator function that takes the value of 1 when the given condition is satisfied.

In terms of the slip and guessing values, bias and root-mean-square error (RMSE) were used for examining their parameter recovery. The bias and RMSE for the $j$-th item guessing value estimates $\hat{g}_{mj}$ were evaluated as
\begin{align*}
Bias_j &= \frac{1}{100} \sum_{m=1}^{100} \left( \hat{g}_{mj} - g^{true}_j \right),\\
RMSE_j &= \sqrt{ \frac{1}{100} \sum_{m=1}^{100} \left( \hat{g}_{mj} - g^{true}_j \right)^2}.
\end{align*}
Finally, we averaged EACR, PACR, bias, and RMSE across the relevant attributes and items. 

\subsection{Settings for Estimation}

For the ML estimation, we employed the ``GDINA" function in the GDINA package \parencite{ma_gdina_2020}. This function estimates DCMs' parameters using the marginal maximum likelihood method with the expectation-maximization (EM) algorithm. The EM iteration was stopped when the maximum absolute difference between consecutive estimates became less than $10^{-4}$ or the number of iterations reached 2000.

Regarding the Bayesian estimation with a Markov chain Monte Carlo (MCMC) method, we utilized the just another Gibbs sampler (JAGS) software program \parencite{plummer_jags_2003} on the R2jags package \parencite{su_r2jags_2020}. To diminish the impact of prior information on parameter estimates, we placed the uniform or weakly-informed priors to the parameters of DCMs. Specifically, for the structural parameter of the five DCMs, $\mathrm{Dir}(\bm{1}_L)$ was assigned, and for their item parameters, $s_j \sim \mathrm{Beta}(1,1)$, $g_j \sim \mathrm{Beta}(1,1)$, $\pi_j \sim \mathrm{Beta}(1,1)$, $r_j \sim \mathrm{Beta}(1,1)$, the main effects of attributes $\lambda_{jk} \sim \mathrm{TruncatedNormal}_{\lambda_{jk}>0}(0,10)$, and the intercept and interaction effects of attributes $\lambda_{j \cdot \cdot} \sim \mathrm{N}(0,10)$ were assigned. We ran three chains of 5000 iterations with 2000 burn-in to construct the posteriors of the five DCMs' parameters and computed the maximum a posteriori (MAP) estimates of attributes and the expected a posteriori (EAP) estimates of item parameters. All the JAGS code for the Bayesian estimation were written based on \textcite{zhan_using_2019}.

Lastly, the nonparametric estimation was conducted by the ``AlphaNP" function in the NPCD package \parencite{zheng_npcd_2019}. Because the nonparametric respondent classification under the noncompensatory and compensatory assumptions are available in this function, we implemented the nonparametric method only for the DINA model, DINO model, RRUM, and CRUM. It should be noted that although we applied this method to the RRUM and CRUM, their results should be taken only as a reference because the RRUM and CRUM do not exhibit the complete noncompensatory and compensatory assumptions. All the R code implemented in this study can be obtained from the Open Science Framework: \url{https://osf.io/hqu5k/?view_only=c07962933c0a4f86803d43afbb8dfb6d}.

\section{Results}
In this section, we first report key findings of respondent classification. Subsequently, we report the results on item parameter estimation. The Online Appendix is available on the Open Science Framework: \url{https://osf.io/hqu5k/?view_only=c07962933c0a4f86803d43afbb8dfb6d}.

Regarding the occurrence of datasets with perfect response patterns, we observed that the maximum number of times this type of dataset was generated in one condition was 29 under the DINA model. Since ML estimation cannot be performed with such datasets, we omitted them in calculating the measures of accuracy for not only the ML but also the Bayesian and nonparametric methods. 

\subsection{Respondent Classification}

For ease of interpretation, we selected item discrimination and the number of items, both of which are known to have a large impact on respondent classification \parencite{paulsen_examining_2019}, as the main factors on which we focus. Then, we calculated the marginal element- and pattern-wise attribute classification rate for each of the other manipulated factors. We present the results of our key findings in the form of tables owing to space limitations. The details of all the results on respondent classification are attached in Tables A1 to A6 in the Online Appendix.

We inspected the measures of respondent classification in the conditions where a salient trend was observed. Similar to the previous study \parencite{paulsen_examining_2019}, item discrimination and the number of items presented a strong influence on respondent classification. When item discrimination was high and $J=40$, respondent classification generally achieved over 90\% and 80\% of the EACR and PACR, where these measures incremented by approximately 25 and 60 percentage points, respectively (for example, see Table A2 in the Online Appendix), compared with the low item discrimination condition. This result illustrates that, given items of high quality, we can estimate respondents' attributes precisely even under a small sample size. Similarly, the increase in the number of items positively contributed to the accuracy of respondent classification, and the PACR under $J=40$ was much larger than under $J=20$ (for example, see Tables A1 and A2 in the Online Appendix). However, the degree of this effect was not as large as that of item discrimination.

Another manipulated factor that showed a notable impact on respondent classification is Q-matrix misspecification. Tables \ref{tab:Table2} and \ref{tab:Table3} represent the marginal element- and pattern-wise attribute classification rate given Q-matrix misspecification. The effect of Q-matrix misspecification differed between EACR and PACR. While the effect of Q-matrix misspecification for EACR was relatively small, its effect for PACR was significantly large. In particular, the PACR under the high item discrimination condition declined considerably as the degree of Q-matrix misspecification increased. The negative effect for the EACR and PACR under the low item discrimination condition was not salient, unlike the ones under the high item discrimination condition. Because the items with low discriminating power themselves are
an indicator of Q-matrix misspeciﬁcation in applications \parencite{rupp_effects_2008}, the Q-matrix misspeciﬁcation manipulated in these conditions may not exert a large negative effect on respondent classiﬁcation. In terms of sample size, the PACR improved as sample size increased (see Tables A1 and A2 in the Online Appendix). The effect of the number of attributes was observed in the PACR, where its value under $K=5$ is smaller than under $K=4$ (see Tables A3 and A4 in the Online Appendix).

We then computed the mean differences of EACR and PACR among the ML, Bayesian, and nonparametric estimation methods. The results in Table \ref{tab:Table4} show clear patterns. The EACR and PACR from the Bayesian method outperformed those from the ML method considering the parsimonious models, such as the DINA and DINO models. Meanwhile, this superiority was reversed in the complex models, such as the RRUM, CRUM, and LCDM. The nonparametric method under the DINA and DINO models---which hold the complete noncompensatory and compensatory assumptions, respectively---demonstrated better performance than both the ML and Bayesian methods. Nonetheless, the differences in EACR and PACR among these three methods were not noticeably large, indicating that these estimation methods would perform comparably in small-scale settings.

\begin{table}[htbp]
 \caption{Marginal EACR and PACR given Q-matrix misspecification when $J=20$}
 \centering
\resizebox{\linewidth}{!}{
\begin{tabular}{rlcccccccc}
\toprule 
& & \multicolumn{8}{c}{Q-matrix Misspecification} \\
\cmidrule{3-10}\multicolumn{2}{c}{\begin{tabular}{c} Item Discrimination: \\ High \end{tabular}} & \multicolumn{2}{c}{0\%} & & \multicolumn{2}{c}{10\%} & & \multicolumn{2}{c}{20\%} \\
\cmidrule{3-4}\cmidrule{6-7}\cmidrule{9-10}\multicolumn{2}{c}{$J=20$} & EACR& PACR& & EACR& PACR& & EACR& PACR \\
\cmidrule{1-4}\cmidrule{6-7}\cmidrule{9-10}
\multicolumn{1}{l}{ML} & DINA& 0.959& 0.856& & 0.935& 0.768& & 0.879& 0.581\\
& DINO& 0.953& 0.839& & 0.918& 0.715& & 0.882& 0.599\\
& RRUM& 0.948& 0.795& & 0.934& 0.745& & 0.906& 0.649\\
& CRUM& 0.950& 0.796& & 0.927& 0.709& & 0.903& 0.630\\
& LCDM& 0.932& 0.737& & 0.909& 0.660& & 0.886& 0.591\\
\midrule
\multicolumn{1}{l}{Bayes} & DINA& 0.959& 0.849& & 0.930& 0.743& & 0.857& 0.507\\
& DINO& 0.949& 0.821& & 0.904& 0.670& & 0.860& 0.522\\
& RRUM& 0.930& 0.718& & 0.910& 0.651& & 0.877& 0.545\\
& CRUM& 0.938& 0.752& & 0.913& 0.660& & 0.887& 0.578\\
& LCDM& 0.923& 0.700& & 0.899& 0.616& & 0.873& 0.536\\
\midrule
\multicolumn{1}{l}{Nonparametric} & DINA& 0.952& 0.840& & 0.927& 0.744& & 0.868& 0.557\\
& DINO& 0.952& 0.835& & 0.906& 0.679& & 0.870& 0.564\\
& RRUM& 0.886& 0.566& & 0.877& 0.544& & 0.856& 0.489\\
& CRUM& 0.872& 0.508& & 0.866& 0.502& & 0.856& 0.488\\
\bottomrule \\
\end{tabular}}
\resizebox{\linewidth}{!}{
\centering
\begin{tabular}{rlcccccccc}
\toprule
& & \multicolumn{8}{c}{Q-matrix Misspecification} \\
\cmidrule{3-10}\multicolumn{2}{c}{\begin{tabular}{c} Item Discrimination: \\ Low \end{tabular}} & \multicolumn{2}{c}{0\%} & & \multicolumn{2}{c}{10\%} & & \multicolumn{2}{c}{20\%} \\
\cmidrule{3-4}\cmidrule{6-7}\cmidrule{9-10}\multicolumn{2}{c}{$J=20$} & EACR& PACR& & EACR& PACR& & EACR& PACR \\
\cmidrule{1-4}\cmidrule{6-7}\cmidrule{9-10}
\multicolumn{1}{l}{ML} & DINA& 0.681& 0.234& & 0.670& 0.205& & 0.653& 0.177\\
& DINO& 0.675& 0.224& & 0.656& 0.189& & 0.652& 0.177\\
& RRUM& 0.667& 0.160& & 0.660& 0.156& & 0.647& 0.146\\
& CRUM& 0.661& 0.157& & 0.655& 0.150& & 0.646& 0.143\\
& LCDM& 0.666& 0.162& & 0.658& 0.155& & 0.649& 0.149\\
\midrule
\multicolumn{1}{l}{Bayes} & DINA& 0.707& 0.273& & 0.690& 0.240& & 0.672& 0.208\\
& DINO& 0.706& 0.256& & 0.683& 0.221& & 0.669& 0.194\\
& RRUM& 0.668& 0.176& & 0.657& 0.164& & 0.644& 0.152\\
& CRUM& 0.655& 0.165& & 0.644& 0.155& & 0.631& 0.139\\
& LCDM& 0.642& 0.160& & 0.634& 0.151& & 0.628& 0.148\\
\midrule
\multicolumn{1}{l}{Nonparametric} & DINA& 0.711& 0.303& & 0.697& 0.260& & 0.678& 0.217\\
& DINO& 0.714& 0.302& & 0.690& 0.251& & 0.681& 0.223\\
& RRUM& 0.688& 0.199& & 0.683& 0.191& & 0.674& 0.181\\
& CRUM& 0.686& 0.198& & 0.679& 0.191& & 0.674& 0.184\\
\bottomrule 
\end{tabular}}%
 \begin{tablenotes}
 \item \footnotesize{\textit{Note.} $J=$ number of items, EACR = element-wise attribute classification rate, PACR = pattern-wise attribute classification rate.}
 \end{tablenotes}
\label{tab:Table2}%
\end{table}

\begin{table}[htbp]
\centering
\caption{Marginal EACR and PACR given Q-matrix misspecification when $J=40$}
\resizebox{\linewidth}{!}{
\begin{tabular}{rlcccccccc}
\toprule 
& & \multicolumn{8}{c}{Q-matrix Misspecification} \\
\cmidrule{3-10}\multicolumn{2}{c}{\begin{tabular}{c} Item Discrimination: \\ High \end{tabular}} & \multicolumn{2}{c}{0\%} & & \multicolumn{2}{c}{10\%} & & \multicolumn{2}{c}{20\%} \\
\cmidrule{3-4}\cmidrule{6-7}\cmidrule{9-10}\multicolumn{2}{c}{$J=40$} & EACR& PACR& & EACR& PACR& & EACR& PACR \\
\cmidrule{1-4}\cmidrule{6-7}\cmidrule{9-10}
\multicolumn{1}{l}{ML} & DINA& 0.987& 0.950& & 0.972& 0.886& & 0.944& 0.779\\
& DINO& 0.985& 0.940& & 0.963& 0.852& & 0.936& 0.753\\
& RRUM& 0.988& 0.946& & 0.982& 0.920& & 0.965& 0.851\\
& CRUM& 0.994& 0.972& & 0.988& 0.950& & 0.980& 0.914\\
& LCDM& 0.987& 0.944& & 0.978& 0.911& & 0.964& 0.860\\
\midrule
\multicolumn{1}{l}{Bayes} & DINA& 0.988& 0.950& & 0.970& 0.876& & 0.943& 0.765\\
& DINO& 0.985& 0.940& & 0.957& 0.827& & 0.924& 0.708\\
& RRUM& 0.987& 0.943& & 0.980& 0.909& & 0.960& 0.829\\
& CRUM& 0.988& 0.948& & 0.980& 0.912& & 0.964& 0.846\\
& LCDM& 0.977& 0.899& & 0.964& 0.848& & 0.944& 0.773\\
\midrule
\multicolumn{1}{l}{Nonparametric} & DINA& 0.985& 0.942& & 0.961& 0.842& & 0.903& 0.629\\
& DINO& 0.982& 0.929& & 0.951& 0.806& & 0.926& 0.714\\
& RRUM& 0.958& 0.820& & 0.951& 0.794& & 0.921& 0.689\\
& CRUM& 0.869& 0.458& & 0.879& 0.514& & 0.874& 0.512\\
\bottomrule \\
\end{tabular}}
\resizebox{\linewidth}{!}{
\begin{tabular}{rlcccccccc}
\toprule 
& & \multicolumn{8}{c}{Q-matrix Misspecification} \\
\cmidrule{3-10}\multicolumn{2}{c}{\begin{tabular}{c} Item Discrimination: \\ Low \end{tabular}} & \multicolumn{2}{c}{0\%} & & \multicolumn{2}{c}{10\%} & & \multicolumn{2}{c}{20\%} \\
\cmidrule{3-4}\cmidrule{6-7}\cmidrule{9-10}\multicolumn{2}{c}{$J=40$} & EACR& PACR& & EACR& PACR& & EACR& PACR \\
\cmidrule{1-4}\cmidrule{6-7}\cmidrule{9-10}
\multicolumn{1}{l}{ML} & DINA& 0.738& 0.325& & 0.718& 0.273& & 0.693& 0.225\\
& DINO& 0.736& 0.313& & 0.705& 0.256& & 0.693& 0.228\\
& RRUM& 0.732& 0.242& & 0.717& 0.223& & 0.703& 0.209\\
& CRUM& 0.732& 0.243& & 0.717& 0.223& & 0.703& 0.209\\
& LCDM& 0.728& 0.237& & 0.712& 0.215& & 0.701& 0.207\\
\midrule
\multicolumn{1}{l}{Bayes} & DINA& 0.773& 0.387& & 0.746& 0.325& & 0.717& 0.265\\
& DINO& 0.773& 0.374& & 0.737& 0.303& & 0.717& 0.266\\
& RRUM& 0.691& 0.202& & 0.670& 0.180& & 0.650& 0.160\\
& CRUM& 0.676& 0.189& & 0.657& 0.172& & 0.637& 0.155\\
& LCDM& 0.650& 0.164& & 0.641& 0.158& & 0.626& 0.145\\
\midrule
\multicolumn{1}{l}{Nonparametric} & DINA& 0.785& 0.431& & 0.761& 0.362& & 0.732& 0.289\\
& DINO& 0.779& 0.411& & 0.748& 0.336& & 0.732& 0.297\\
& RRUM& 0.733& 0.250& & 0.722& 0.234& & 0.708& 0.220\\
& CRUM& 0.738& 0.256& & 0.732& 0.255& & 0.723& 0.245\\
\bottomrule 
\end{tabular}}%
 \begin{tablenotes}
 \item \footnotesize{\textit{Note.} $J=$ number of items, EACR = element-wise attribute classification rate, PACR = pattern-wise attribute classification rate.}
 \end{tablenotes}
\label{tab:Table3}%
\end{table}

\begin{table}[htbp]
\centering
\caption{Mean difference of EACR and PACR among the ML, Bayes, and nonparametric estimations}
\begin{tabular}{rlccccc}
\toprule 
& & \multicolumn{5}{c}{Models} \\
\cmidrule{3-7}& & DINA& DINO& RRUM& CRUM& LCDM \\
\midrule
    \multicolumn{1}{l}{EACR} & Bayes - ML & 0.010  & 0.009  & -0.019  & -0.024  & -0.031  \\
          & NP - ML & 0.011  & 0.015  & -0.016  & -0.034  & - \\
          & NP - Bayes & 0.001  & 0.006  & 0.003  & -0.010  & - \\
    \midrule
    \multicolumn{1}{l}{PACR} & Bayes - ML & 0.011  & 0.002  & -0.034  & -0.035  & -0.044  \\
          & NP - ML & 0.013  & 0.022  & -0.072  & -0.149  & - \\
          & NP - Bayes & 0.005  & 0.021  & -0.036  & -0.112  & - \\
\bottomrule 
\end{tabular}%
 \begin{tablenotes}
 \item \footnotesize{\textit{Note.} NP $=$ Nonparametric, EACR = element-wise attribute classification rate, PACR = pattern-wise attribute classification rate.}
 \end{tablenotes}
\label{tab:Table4}%
\end{table}%

\subsection{Item Parameter Estimation}

In the same manner as with respondent classification, we selected item discrimination and sample size as the main factors on which we focus. Subsequently, we calculated the marginal bias and RMSE for each of the other manipulated factors. We present the results of key findings in the form of tables owing to space limitations. The details of all the results on item parameter estimation are attached in Tables A7 to A18 in the Online Appendix.

Overall, the factor of item discrimination did not have a remarkable impact on the values of bias and RMSE, and only the direction of bias was generally changed according to whether item discrimination was high or low. Specifically, the direction of bias was positive when item discrimination was high and vice versa (for example, see Tables \ref{tab:Table5}, \ref{tab:Table6}, and \ref{tab:Table7}). Contrary to item discrimination, the factor of sample size was influential in item parameter estimation (for example, see Tables A7, A8, and A9 in the Online Appendix). The RMSEs of the slip and guessing values were considerably large under the conditions with the sample sizes of 20 and 40. Although the values of the RMSE became smaller under the sample size of 160, the RMSEs were still poor. This result indicates that the item parameter estimates under small sample sizes are not reliable. Hence, these estimates should be taken only as a reference. 

Next, we investigated the results in the conditions where a notable trend was found. Similar to the results in respondent classification, Q-matrix misspecification also had an impact on the accuracy of item parameter estimation. As Q-matrix misspecification worsened, the bias and RMSE of the slip and guessing values increased, which is congruent with previous studies \parencite{rupp_effects_2008,kunina-habenicht_impact_2012}. In particular, its negative effect was more salient under the high item discrimination condition than under the low item discrimination condition. Next, Tables \ref{tab:Table5}, \ref{tab:Table6}, and \ref{tab:Table7} show the bias and RMSE given the number of items. The increase in the number of items generally reduced the bias and RMSE of the slip and guessing values. Lastly, as the number of attributes increased, the bias and RMSE of the slip and guessing values increased (see Tables A10, A11, and A12 in the Online Appendix). This effect will be more prominent when a large $K$ is specified in a Q-matrix.

\begin{table}[htbp]
\centering
\caption{Marginal bias and RMSE of the slip and guessing values given the number of items when $N = 20$}
\resizebox{\linewidth}{!}{
\begin{tabular}{rcccccccccccc}
\toprule 
& & \multicolumn{11}{c}{Number of Items} \\
\cmidrule{3-13}& & \multicolumn{5}{c}{$J=20$}& & \multicolumn{5}{c}{$J=40$} \\
\cmidrule{3-7}\cmidrule{9-13}\multicolumn{2}{c}{\begin{tabular}{c} Item Discrimination: \\ High \end{tabular}} & \multicolumn{2}{c}{slip} & & \multicolumn{2}{c}{guessing} & & \multicolumn{2}{c}{slip} & & \multicolumn{2}{c}{guessing} \\
\cmidrule{3-4}\cmidrule{6-7}\cmidrule{9-10}\cmidrule{12-13}\multicolumn{2}{c}{$N=20$} & Bias& RMSE& & Bias& RMSE& & Bias& RMSE& & Bias& RMSE \\
\midrule
\multicolumn{1}{l}{ML} & DINA& 0.053& 0.193& & 0.031& 0.114& & 0.039& 0.178& & 0.036& 0.101\\
& DINO& 0.034& 0.116& & 0.037& 0.179& & 0.028& 0.109& & 0.046& 0.167\\
& RRUM& 0.047& 0.179& & 0.040& 0.149& & 0.035& 0.178& & 0.047& 0.134\\
& CRUM& 0.037& 0.151& & 0.032& 0.166& & 0.029& 0.154& & 0.040& 0.152\\
& LCDM& 0.043& 0.177& & 0.026& 0.182& & 0.032& 0.167& & 0.045& 0.171\\
\midrule
\multicolumn{1}{l}{Bayes} & DINA& 0.244& 0.277& & 0.086& 0.119& & 0.210& 0.248& & 0.089& 0.114\\
& DINO& 0.116& 0.144& & 0.171& 0.204& & 0.098& 0.132& & 0.171& 0.201\\
& RRUM& 0.200& 0.239& & 0.068& 0.115& & 0.169& 0.214& & 0.078& 0.113\\
& CRUM& 0.005& 0.102& & 0.093& 0.177& & 0.007& 0.113& & 0.080& 0.152\\
& LCDM& 0.012& 0.143& & 0.110& 0.190& & 0.004& 0.144& & 0.118& 0.189\\
\bottomrule \\
\end{tabular}}%

\resizebox{\linewidth}{!}{
\begin{tabular}{rcccccccccccc}
\toprule 
& & \multicolumn{11}{c}{Number of Items} \\
\cmidrule{3-13}& & \multicolumn{5}{c}{$J=20$}& & \multicolumn{5}{c}{$J=40$} \\
\cmidrule{3-7}\cmidrule{9-13}\multicolumn{2}{c}{\begin{tabular}{c} Item Discrimination: \\ Low \end{tabular}} & \multicolumn{2}{c}{slip} & & \multicolumn{2}{c}{guessing} & & \multicolumn{2}{c}{slip} & & \multicolumn{2}{c}{guessing} \\
\cmidrule{3-4}\cmidrule{6-7}\cmidrule{9-10}\cmidrule{12-13}\multicolumn{2}{c}{$N=20$} & Bias& RMSE& & Bias& RMSE& & Bias& RMSE& & Bias& RMSE \\
\midrule
\multicolumn{1}{l}{ML} & DINA& -0.012& 0.350& & -0.017& 0.179& & 0.010& 0.337& & 0.015& 0.162\\
& DINO& -0.016& 0.184& & -0.043& 0.329& & 0.011& 0.161& & -0.029& 0.323\\
& RRUM& -0.049& 0.276& & -0.031& 0.248& & -0.028& 0.253& & -0.006& 0.227\\
& CRUM& -0.056& 0.278& & -0.043& 0.275& & -0.028& 0.246& & -0.021& 0.251\\
& LCDM& -0.065& 0.309& & -0.065& 0.308& & -0.040& 0.278& & -0.030& 0.295\\
\midrule
\multicolumn{1}{l}{Bayes} & DINA& 0.063& 0.135& & -0.011& 0.106& & 0.066& 0.141& & -0.017& 0.108\\
& DINO& 0.033& 0.122& & -0.002& 0.103& & 0.031& 0.118& & -0.009& 0.110\\
& RRUM& 0.012& 0.131& & -0.107& 0.145& & 0.055& 0.140& & -0.143& 0.171\\
& CRUM& -0.213& 0.251& & -0.083& 0.171& & -0.238& 0.262& & -0.013& 0.159\\
& LCDM& -0.106& 0.275& & -0.071& 0.168& & -0.128& 0.273& & -0.016& 0.162\\
\bottomrule 
\end{tabular}}%
 \begin{tablenotes}
 \item \footnotesize{\textit{Note.} $N$ = sample size, $J$ = number of items, RMSE = root-mean-square error.}
 \end{tablenotes}
\label{tab:Table5}%
\end{table}%

\begin{table}[htbp]
\centering
\caption{Marginal bias and RMSE of the slip and guessing values given the number of items when $N = 40$}
\resizebox{\linewidth}{!}{
\begin{tabular}{rcccccccccccc}
\toprule 
& & \multicolumn{11}{c}{Number of Items} \\
\cmidrule{3-13}& & \multicolumn{5}{c}{$J=20$}& & \multicolumn{5}{c}{$J=40$} \\
\cmidrule{3-7}\cmidrule{9-13}\multicolumn{2}{c}{\begin{tabular}{c} Item Discrimination: \\ High \end{tabular}} & \multicolumn{2}{c}{slip} & & \multicolumn{2}{c}{guessing} & & \multicolumn{2}{c}{slip} & & \multicolumn{2}{c}{guessing} \\
\cmidrule{3-4}\cmidrule{6-7}\cmidrule{9-10}\cmidrule{12-13}\multicolumn{2}{c}{$N=40$} & Bias& RMSE& & Bias& RMSE& & Bias& RMSE& & Bias& RMSE \\
\midrule
\multicolumn{1}{l}{ML} & DINA& 0.040& 0.147& & 0.033& 0.095& & 0.042& 0.148& & 0.033& 0.082\\
& DINO& 0.031& 0.095& & 0.045& 0.159& & 0.027& 0.085& & 0.050& 0.146\\
& RRUM& 0.040& 0.140& & 0.037& 0.122& & 0.029& 0.132& & 0.036& 0.100\\
& CRUM& 0.039& 0.124& & 0.025& 0.127& & 0.022& 0.113& & 0.031& 0.109\\
& LCDM& 0.044& 0.138& & 0.026& 0.145& & 0.027& 0.124& & 0.035& 0.125\\
\midrule
\multicolumn{1}{l}{Bayes} & DINA& 0.174& 0.214& & 0.063& 0.096& & 0.145& 0.190& & 0.063& 0.088\\
& DINO& 0.077& 0.107& & 0.138& 0.181& & 0.064& 0.096& & 0.137& 0.175\\
& RRUM& 0.149& 0.189& & 0.056& 0.105& & 0.112& 0.157& & 0.057& 0.093\\
& CRUM& 0.010& 0.096& & 0.057& 0.131& & 0.010& 0.096& & 0.044& 0.103\\
& LCDM& 0.002& 0.110& & 0.096& 0.167& & 0.003& 0.113& & 0.074& 0.133\\
\bottomrule \\
\end{tabular}}%

\resizebox{\linewidth}{!}{
\begin{tabular}{rcccccccccccc}
\toprule 
& & \multicolumn{11}{c}{Number of Items} \\
\cmidrule{3-13}& & \multicolumn{5}{c}{$J=20$}& & \multicolumn{5}{c}{$J=40$} \\
\cmidrule{3-7}\cmidrule{9-13}\multicolumn{2}{c}{\begin{tabular}{c} Item Discrimination: \\ Low \end{tabular}} & \multicolumn{2}{c}{slip} & & \multicolumn{2}{c}{guessing} & & \multicolumn{2}{c}{slip} & & \multicolumn{2}{c}{guessing} \\
\cmidrule{3-4}\cmidrule{6-7}\cmidrule{9-10}\cmidrule{12-13}\multicolumn{2}{c}{$N=40$} & Bias& RMSE& & Bias& RMSE& & Bias& RMSE& & Bias& RMSE \\
\midrule
\multicolumn{1}{l}{ML} & DINA& -0.032& 0.292& & -0.009& 0.149& & -0.016& 0.259& & 0.017& 0.131\\
& DINO& -0.005& 0.156& & -0.032& 0.284& & 0.013& 0.128& & -0.021& 0.265\\
& RRUM& -0.035& 0.244& & -0.023& 0.213& & -0.016& 0.216& & 0.000& 0.184\\
& CRUM& -0.037& 0.239& & -0.027& 0.239& & -0.014& 0.205& & -0.001& 0.211\\
& LCDM& -0.053& 0.268& & -0.050& 0.269& & -0.035& 0.230& & -0.027& 0.241\\
\midrule
\multicolumn{1}{l}{Bayes} & DINA& 0.049& 0.121& & -0.002& 0.091& & 0.051& 0.123& & -0.005& 0.091\\
& DINO& 0.027& 0.103& & 0.002& 0.101& & 0.024& 0.096& & 0.000& 0.108\\
& RRUM& 0.001& 0.121& & -0.080& 0.128& & 0.045& 0.120& & -0.120& 0.153\\
& CRUM& -0.184& 0.227& & -0.078& 0.155& & -0.222& 0.249& & -0.013& 0.140\\
& LCDM& -0.086& 0.265& & -0.039& 0.142& & -0.128& 0.267& & 0.026& 0.136\\
\bottomrule 
\end{tabular}}%
 \begin{tablenotes}
 \item \footnotesize{\textit{Note.} $N$ = sample size, $J$ = number of items, RMSE = root-mean-square error.}
 \end{tablenotes}
\label{tab:Table6}%
\end{table}%

\begin{table}[htbp]
\centering
\caption{Marginal bias and RMSE of the slip and guessing values given the number of items when $N = 160$}
\resizebox{\linewidth}{!}{
\begin{tabular}{rcccccccccccc}
\toprule 
& & \multicolumn{11}{c}{Number of Items} \\
\cmidrule{3-13}& & \multicolumn{5}{c}{$J=20$}& & \multicolumn{5}{c}{$J=40$} \\
\cmidrule{3-7}\cmidrule{9-13}\multicolumn{2}{c}{\begin{tabular}{c} Item Discrimination: \\ High \end{tabular}} & \multicolumn{2}{c}{slip} & & \multicolumn{2}{c}{guessing} & & \multicolumn{2}{c}{slip} & & \multicolumn{2}{c}{guessing} \\
\cmidrule{3-4}\cmidrule{6-7}\cmidrule{9-10}\cmidrule{12-13}\multicolumn{2}{c}{$N=160$} & Bias& RMSE& & Bias& RMSE& & Bias& RMSE& & Bias& RMSE \\
\midrule
\multicolumn{1}{l}{ML} & DINA& 0.036& 0.102& & 0.036& 0.073& & 0.045& 0.112& & 0.032& 0.063\\
& DINO& 0.036& 0.073& & 0.040& 0.109& & 0.027& 0.061& & 0.049& 0.115\\
& RRUM& 0.031& 0.091& & 0.030& 0.080& & 0.031& 0.091& & 0.022& 0.066\\
& CRUM& 0.034& 0.089& & 0.024& 0.084& & 0.022& 0.074& & 0.026& 0.078\\
& LCDM& 0.040& 0.095& & 0.023& 0.088& & 0.021& 0.073& & 0.031& 0.084\\
\midrule
\multicolumn{1}{l}{Bayes} & DINA& 0.085& 0.127& & 0.046& 0.072& & 0.075& 0.123& & 0.041& 0.064\\
& DINO& 0.052& 0.078& & 0.070& 0.115& & 0.038& 0.064& & 0.074& 0.121\\
& RRUM& 0.066& 0.105& & 0.038& 0.078& & 0.053& 0.097& & 0.031& 0.064\\
& CRUM& 0.024& 0.083& & 0.028& 0.085& & 0.016& 0.072& & 0.030& 0.078\\
& LCDM& 0.015& 0.085& & 0.044& 0.095& & 0.012& 0.073& & 0.040& 0.083\\
\bottomrule\\
\end{tabular}}

\resizebox{\linewidth}{!}{
\begin{tabular}{rcccccccccccc}
\toprule 
& & \multicolumn{11}{c}{Number of Items} \\
\cmidrule{3-13}& & \multicolumn{5}{c}{$J=20$}& & \multicolumn{5}{c}{$J=40$} \\
\cmidrule{3-7}\cmidrule{9-13}\multicolumn{2}{c}{\begin{tabular}{c} Item Discrimination: \\ Low \end{tabular}} & \multicolumn{2}{c}{slip} & & \multicolumn{2}{c}{guessing} & & \multicolumn{2}{c}{slip} & & \multicolumn{2}{c}{guessing} \\
\cmidrule{3-4}\cmidrule{6-7}\cmidrule{9-10}\cmidrule{12-13}\multicolumn{2}{c}{$N=160$} & Bias& RMSE& & Bias& RMSE& & Bias& RMSE& & Bias& RMSE \\
\midrule
\multicolumn{1}{l}{ML} & DINA& -0.015& 0.175& & -0.006& 0.105& & 0.010& 0.127& & 0.007& 0.083\\
& DINO& -0.004& 0.114& & -0.009& 0.180& & 0.007& 0.081& & 0.009& 0.132\\
& RRUM& -0.005& 0.182& & -0.004& 0.150& & 0.004& 0.142& & 0.018& 0.118\\
& CRUM& -0.014& 0.179& & -0.005& 0.180& & 0.005& 0.135& & 0.013& 0.136\\
& LCDM& -0.012& 0.208& & -0.009& 0.207& & 0.002& 0.165& & 0.005& 0.162\\
\midrule
\multicolumn{1}{l}{Bayes} & DINA& 0.041& 0.109& & 0.006& 0.071& & 0.043& 0.101& & 0.008& 0.065\\
& DINO& 0.021& 0.078& & 0.013& 0.102& & 0.020& 0.066& & 0.017& 0.096\\
& RRUM& 0.008& 0.103& & -0.037& 0.097& & 0.037& 0.094& & -0.064& 0.108\\
& CRUM& -0.117& 0.176& & -0.046& 0.124& & -0.138& 0.185& & -0.019& 0.113\\
& LCDM& -0.064& 0.245& & 0.024& 0.109& & -0.102& 0.257& & 0.079& 0.117\\
\bottomrule 
\end{tabular}}%
 \begin{tablenotes}
 \item \footnotesize{\textit{Note.} $N$ = sample size, $J$ = number of items, RMSE = root-mean-square error.}
 \end{tablenotes}
\label{tab:Table7}%
\end{table}%

Regarding the difference between the ML and Bayesian methods, we found that the RMSE from the ML estimates for the slip and guessing values under the low item discrimination condition were approximately two times larger than under the high item discrimination condition. Additionally, under the low item discrimination condition, as the complexity of DCMs increased, the bias and RMSE from the EAP estimates for them deteriorated, and the reverse pattern was observed under the high item discrimination condition. Moreover, the RMSE of the ML estimates for the slip and guessing values were significantly larger than that of the EAP estimates for them under the low item discrimination condition. These tendencies of the ML and Bayesian methods became less salient when the sample size was 160. For examples of the aforementioned differences, see Tables \ref{tab:Table5}, \ref{tab:Table6}, and \ref{tab:Table7}.

It is noteworthy that the bias and RMSE of the ML estimates for the slip values of the DINA model and RRUM and for the guessing values of the DINO model under the high item discrimination condition were significantly different (that is, smaller) from those of the EAP estimate (for example, see Table \ref{tab:Table5}). The bias and RMSE of the EAP estimates for the slip values of the CRUM and LCDM under the low item discrimination condition were much larger than those of the ML estimates.

The former asymmetry with regard to the ML method may entail the occurrence of the boundary problem \parencite{philipp_estimation_2018,maris_estimating_1999}, where the ML estimates for item parameters converge to the bound of their support. In the case of the DINA and DINO models, the boundary problem means that the ML estimates for the slip and guessing parameters converge to 0 or 1. It is known that this problem frequently occurs when there are one or more latent classes to which none of the respondents are deemed to belong, which often occurs in small sample contexts \parencite{decarlo_analysis_2011}. Since DCMs can be framed as a restricted latent class model \parencite{rupp_unique_2008}, the increase in the number of attributes leads to the exponential expansion of possible attribute mastery profiles, which correspond to latent classes in the latent class model.

The latter asymmetry with regard to the Bayesian method is caused by the multimodal posterior of slip values. With further inspection, we observed that the posterior variance of the item parameters for the CRUM and LCDM were particularly large in the small sample size. This large variance leads to another problem when summarizing the posterior of the slip values. Because the CRUM and LCDM have a binomial likelihood with a logistic link function, converting their item parameters to the corresponding slip values involves non-linear transformation. Specifically, their slip values tend to be extreme such as 0 or 1, when an MCMC sample set for the CRUM and LCDM item parameters becomes small or large at the same time. This phenomenon will be more salient as the posterior variance of their item parameters becomes larger. Hence, the multimodality in the posterior of slip values is a problem in the Bayesian method when estimating the slip values of the DCMs with a logistic link under small sample contexts. It should be noted that since the calculation of the guessing values involves only the intercept parameters, the effect of their large posterior variance does not appear.

These observations suggest that the ML and Bayesian methods may perform poorly when estimating the slip and guessing values under small sample conditions.

\section{Discussion and Conclusion}
In this study, we conducted a far-reaching examination on the DINA model, DINO model, RRUM, CRUM, and LCDM with the ML, Bayesian, and nonparametric estimation methods under a simulation design that resembles classroom assessments. We briefly discuss the contributions of this study and then summarize its findings. 

This study made three distinct contributions in comparison with previous simulation studies pertinent to small sample sizes. First and foremost, this study considered different estimation methods for assessing the performance of DCMs under small sample conditions. To gain knowledge of item property, we often appeal to parametric methods such as the ML estimation to estimate item parameters that capture the characteristic of used items. Nevertheless, the frequent occurrence of an item with a perfect response pattern in classroom sample sizes renders the ML estimation infeasible in practice. Our simulation study addressed this limitation by including Bayesian and nonparametric estimation methods and examined the behavior of DCMs' estimates under various conditions that mimic classroom assessments. Second, we unveiled some problems that are present particularly in classroom sample sizes such as 20 and 40. Specifically, the Bayesian method exhibited instability in estimating slip values because of the multimodality of their posteriors under complex DCMs, and the ML method provided irregular item parameter estimates because of a boundary problem under parsimonious DCMs. These findings inform practitioners which of the Bayesian and ML methods can be appropriately used under a user-specified model when they estimate item parameters. Third, although several factors in this simulation, such as Q-matrix misspecification and item discrimination, were also considered in the preceding studies, our simulation design put the main focus on realistic classroom sample sizes such as 20 and 40 and investigated how these factors influenced the estimates of DCMs' parameters in such sample sizes. As applied researchers have recently conceived a practical framework to integrate diagnostic assessments into classroom instruction \parencite{fan_integrating_2021}, the results from this classroom-related simulation study could help such practitioners implement DCMs in an appropriate manner.

Here, we summarize the general findings of this study. For both respondent classification and item parameter estimation, Q-matrix misspecification had a strong impact on estimation accuracy. In particular, when item discrimination was high, the negative effect of Q-matrix misspecification became salient. To avoid the use of a misspecified Q-matrix, we recommend creating some competing Q-matrices and evaluating them with relative model fit measures such as the widely applicable information criterion \parencite[WAIC;][]{watanabe_asymptotic_2010} and widely applicable Bayesian information criterion \parencite[WBIC;][]{watanabe_widely_2013}, in addition to carefully considering each item and its associated attributes. Alternatively, Q-matrix estimation or validation methods can be utilized to detect Q-matrix misspecification. The recent developments in these methods provide practitioners with the tools to infer or validate the elements of a Q-matrix based on likelihood or item discrimination \parencite[e.g.,][]{de_la_torre_general_2016,wang_em-based_2018,xu_identifying_2018,chung_gibbs_2019,liu_constrained_2020}. Since a Q-matrix determines the loading matrix between items and attributes, its statistical inference will yield an estimated Q-matrix that generally fits the observed data better than a predefined Q-matrix.

Considering respondent classification, item discrimination had a marked impact on its accuracy. When the slip and guessing values ranged from 0 to 0.15, the EACR and PACR of the three estimation methods generally attained over 90\% and 80\%, respectively. Meanwhile, their values decreased to approximately $65\% \sim 75\%$ and $20\% \sim 35\%$ under the low item discrimination condition where their true values ranged from 0.25 to 0.4. Hence, items of high quality are required for reliable respondent classification especially in a classroom assessment. Additionally, the number of items moderately contributed to the accuracy of respondent classification. If cross-cutting attributes can be designated in different classroom assessments, concatenating them to increase the number of items in a Q-matrix can be a possible means of classifying respondents more accurately. Regarding the differences in EACR and PACR among the three estimation methods, we found that while the Bayesian method performs better in the parsimonious DCMs than the ML method, in the complex DCMs, the ML method performs better than the Bayesian method. The nonparametric method showed the best accuracy among the three estimation methods in terms of the DINA and DINO models. However, the differences in EACR and PACR among these three methods were negligible.

With respect to item parameter estimation, the factor of sample size presented a notable effect on item parameter estimation. Specifically, under the sample sizes of 20 and 40, the RMSE of the slip and guessing values became considerably large. In general, a typical sample size in a classroom assessment would be between 20 and 40. Thus, the estimates of item parameters should be considered only as a reference in small-scale applications. Another finding related to item parameter estimation is that the bias and RMSE of the slip and guessing values generally declined as the number of items increased. Because the number of items also showed a positive effect on respondent classification, augmenting the number of items in a classroom assessment will be beneficial to utilizing DCMs with high precision.

Certain noticeable behaviors of the ML estimates for the slip and guessing values in the DINA model, DINO model, and RRUM and of the EAP estimates for them in the CRUM and LCDM, respectively, were found. Specifically, the bias and RMSE of these ML estimates were much smaller than those of the EAP estimates. Indeed, we conducted an additional analysis concerning the DINA and DINO models under the high item discrimination and $N=20$ and found that a large number of the ML estimates for the slip parameters of the DINA model and for the guessing parameters of the DINO model converged to 0.0001, indicating the boundary problem of the latent class model. This problem often occurs in small sample contexts \parencite{decarlo_analysis_2011}. Accordingly, the use of the ML method for item parameter estimation in classroom assessments is likely to produce irregular estimates that only appear to be well-estimated. In contrast, the Bayesian method presented a multimodal posterior for the slip values under the CRUM and LCDM due to non-linear transformation of their item parameters. However, the use of these complex DCMs under small sample conditions may not be preferable for two reasons. The first reason is that respondent classification under the complex DCMs is less accurate than under the parsimonious DCMs in both the ML and Bayesian methods, owing to the stricter sample requirement of these complex models. The second reason is that a previous study found that a sample size of over $N=1000$ is still unsatisfactory for accurate estimation of the interaction effect of attributes \parencite{kunina-habenicht_impact_2012}. Hence, although both the ML and Bayesian methods cannot be recommended for item parameter estimation under a small sample size, the problem of the ML method under the parsimonious models is more serious than that of the Bayesian method in practice.

The limitations of this study are twofold. First, this study applied five DCMs to datasets generated from the corresponding true models. However, the true data-generating process is not known in real applications, and hence, model misspecification can occur. Investigating the performance of DCMs in such situations is a promising direction for future research. Second, although this study only considers binary attributes, polytomous attributes that capture the qualitative ordering of cognitive complexity in an attribute \parencite{tzur_karelitz_ordered_2004} have practical importance. As the number of possible combinations in polytomous attribute mastery profiles expands at a much faster pace than that of dichotomous attribute mastery profiles, the effect of the factors related to attributes will be different. Hence, the polytomous case of this simulation study is needed for a better understanding of DCMs under small sample conditions.

In a nutshell, this simulation study with a complex, albeit comprehensive design advanced our understanding of the behavior of DCMs' parameter estimates in small sample contexts. Based on the results of respondent classification and item parameter estimation, the Bayesian method may be a preferable choice in sample sizes of 20 and 40 when practitioners wish to obtain both respondent and item parameters' estimates. In addition, the Bayesian paradigm offers a way of incorporating prior information into an estimation process \parencite{gelman_bayesian_2013}. This benefit can be exploited particularly in a classroom assessment because teachers have a rich knowledge of their students and subjects. Finally, given good specifications of a Q-matrix and well-developed items, DCMs can provide fruitful information about attribute mastery states for both teachers and students under conditions that resemble a classroom assessment, although the results from DCMs should not be used as the decisive criteria of decision-making in classroom instruction.

\section*{Statements and Declarations}
\subsection*{Competing interests}
The authors declare that they have no conflict of interest.

\subsection*{Funding}
This work was supported by JSPS KAKENHI (Grant Number 19H00616 and 21H00936).

\section*{References}
\printbibliography[heading=none]

\end{document}